\documentclass[useAMS,usenatbib,usegraphicx]{mn2e}

\usepackage{graphicx}
\DeclareGraphicsExtensions{.pdf,.png,.jpg,.ps}
\usepackage{epsfig}

\usepackage[bottom]{footmisc}
\usepackage{color,amsmath,enumitem}
\usepackage{threeparttable}
\usepackage{graphicx}
\DeclareGraphicsExtensions{.pdf,.eps,.jpg,.ps,.png}
\usepackage{epsfig}
\usepackage{url}
\usepackage{ amssymb }
\usepackage{pdflscape}
\usepackage{afterpage}
\usepackage{capt-of}
\usepackage{gensymb}

\newcommand{\LMCdistHB}{18.59 $\pm$ 0.06 }  

\newcommand{\LMCdistSD}{18.40 $\pm$ 0.04 }
\newcommand{\LMCdistSDKpc}{47.94 $\pm$ 0.32 kpc}

\newcommand{\LMCdistDepr}{18.41 $\pm$ 0.04 }
\newcommand{\LMCdistDeprKpc}{48.29 $\pm$ 0.38 kpc}

\newcommand{\LMCavgdist}{18.52 $\pm$ 0.05} 
\newcommand{\LMCavgdistKpc}{50.51 $\pm$ 1.19 kpc} 

\begin{document}

\title[Relative ages and distances for LMC GCs]{Exploring the nature and synchronicity of early cluster formation in the Large Magellanic Cloud: II. Relative ages and distances for six ancient globular clusters}
\author[Wagner-Kaiser et al.] {R.~Wagner-Kaiser$^1$, Dougal~Mackey$^2$, Ata~Sarajedini$^{1, 3}$, Brian~Chaboyer$^{4}$, 
\newauthor{Roger~E.~Cohen$^{5}$, Soung-Chul~Yang$^{6}$, Jeffrey~D.~Cummings$^{7}$, Doug Geisler$^{8}$, }
\newauthor{Aaron~J.~Grocholski$^{9}$}
\\
    $^1$University of Florida, Department of Astronomy, 211 Bryant Space Science Center, Gainesville, FL, 32611 USA\\
    $^2$Australian National University, Research School of Astronomy \& Astrophysics, Canberra, ACT 2611, Australia\\
    $^3$Florida Atlantic University, Department of Physics, 777 Glades Rd, Boca Raton, FL, 33431 USA\\
    $^4$Dartmouth College, Department of Physics and Astronomy, Hanover, NH, 03755, USA\\
    $^5$Space Telescope Science Institute, Baltimore, MD 21218, USA\\ 
    $^6$Korea Astronomy and Space Science Institute (KASI), Daejeon 305-348, Korea \\ 
    $^7$Center for Astrophysical Sciences, Johns Hopkins University, Baltimore MD 21218\\
    $^8$Departamento de Astronomia, Universidad de Concepcion, Casilla 160-C, Concepcion, Chile \\
    $^9$Department of Physics and Astronomy, Swarthmore College, Swarthmore, PA 19081, USA\\ }

\date{}

\pagerange{\pageref{firstpage}--\pageref{lastpage}} \pubyear{2002}

\maketitle

\label{firstpage}


\begin{abstract}
We analyze Hubble Space Telescope observations of six globular clusters in the Large Magellanic Cloud from program GO-14164 in Cycle 23. These are the deepest available observations of the LMC globular cluster population; their uniformity facilitates a precise comparison with globular clusters in the Milky Way. Measuring the magnitude of the main sequence turnoff point relative to template Galactic globular clusters allows the relative ages of the clusters to be determined with a mean precision of 8.4\%, and down to 6\% for individual objects. We find that the mean age of our LMC cluster ensemble is identical to the mean age of the oldest metal-poor clusters in the Milky Way halo to 0.2 $\pm$ 0.4 Gyr. This provides the most sensitive test to date of the synchronicity of the earliest epoch of globular cluster formation in two independent galaxies. Horizontal branch magnitudes and subdwarf fitting to the main sequence allow us to determine distance estimates for each cluster, and examine their geometric distribution in the LMC. Using two different methods, we find an average distance to the LMC of \LMCavgdist.

\end{abstract}



\section{Introduction}

Globular clusters (GCs) have long been a crucial tool to understand star formation processes in the early Universe as well as the buildup of galactic halo populations. In recent years, high-precision photometry from the Hubble Space Telescope (HST) has facilitated accurate and consistent analyses of such clusters. In particular, the ACS Galactic Globular Cluster Treasury Program has provided a deep and uniform photometric database for a large sample of Galactic GCs. Its enduring contribution to the literature includes a wide range of studies of GC properties, such as mass functions, binary fractions, and horizontal branch morphology, among others (\citealt{Dotter:2010,Marin-Franch:2009,Milone:2012,Paust:2010,Sarajedini:2007,Wagner-Kaiser:2017}).

However, to fully untangle the nature and role of GCs in the formation and evolution of galaxies, it is necessary to look beyond the thoroughly-studied Galactic globular cluster (GGC) system. The population of old globular clusters in the Large Magellanic Cloud (LMC) provides the logical next step in expanding the sample of targets for which extremely high quality photometry can be obtained. We have recently obtained very deep, uniform HST observations of six ancient LMC clusters -- NGC 1466, NGC 1841, NGC 2210, NGC 2257, Hodge 11, and Reticulum -- resulting in photometry extending well down their main sequences for the first time (\citealt{Mackey:2017}). We can now begin to explore with unprecedented precision how the timing of the earliest epoch of star cluster formation in the LMC compares to that in the Milky Way, and search for evidence of an age spread and/or a well-defined age-metallicity relationship (AMR) in the ancient LMC cluster system. 

These goals are of particular significance because recent proper motion measurements have demonstrated that the Magellanic system is likely on its first passage about the Milky Way (e.g., \citealt{Kallivayalil:2006,Kallivayalil:2013,Besla:2007}). The implication is that at the time of globular cluster formation the two galaxies were widely separated. Moreover, the large difference in halo mass and observed star-formation histories between the LMC and the Milky Way suggests that their internal environments are likely to have been substantially different from each other. Thus, examining the relative ages of the oldest GCs in the LMC and the Milky Way allows us to begin to quantitatively probe how the earliest epochs of star formation might vary as a function of location and environment.

The past several decades have seen a strong push to better understand the chronology of globular cluster formation in the LMC. \cite{Brocato:1996} studied NGC 1786, NGC 1841 and NGC 2210 and determined ages to $\sim 20\%$ precision, concluding that metal-poor clusters in the LMC were coeval with those in the Milky Way to within $\sim 3$ Gyr. Six additional LMC clusters were studied by \cite{Olsen:1998}, who observed a similar consistency with GGC halo clusters in terms of age, metallicity, and HB morphology. Further work by \cite{Johnson:1999} and \cite{Mackey:2004,Mackey:2006} demonstrated that metal-poor clusters in the LMC and Milky Way are coeval within $\sim 1.5-2$ Gyr. In comparison, \cite{Harris:1997} showed that the earliest GCs in the Galactic halo, out to a radius of some 100 kpc (well beyond the current distance of the LMC) were all formed within 1 Gyr.

In this paper we determine the ages of the six clusters in our LMC sample relative to metal-poor GCs in the Milky Way with substantially greater precision than any previous work. While absolute ages in principle allow us to place strong constraints on the age of the Universe, in practice such measurements are significantly less precise than relative age measurements due to persistent systematic uncertainties (e.g.: metallicity measurements, particular aspects of stellar evolution, foreground reddening, etc.) such that relative ages are preferred for this type of analysis. One key advantage that we exploit here is the availability of data from the Galactic Globular Cluster Treasury Program, obtained using the same telescope, camera and filters as our LMC observations, and to a comparable depth. By following closely the methodology developed by \cite{Marin-Franch:2009}, who explored the relative ages of clusters in the Treasury sample, we measure relative ages for our LMC clusters on the same scale and as free from biases and systematics as possible. 

In addition to the above, we also use our new data to obtain a precise cluster-based distance estimate to the LMC. Because our photometry extends several magnitudes down the main sequence for each target, we are able to employ the method of sub-dwarf fitting utilising the high quality parallax measurements for several nearby main sequence stars determined by \cite{Chaboyer:2017} (GO-11704 and GO-12320). We compare our distance measurements to those derived from the luminosity of the horizontal branch (HB) -- a region of the colour-magnitude diagram observed with very high signal-to-noise ($\approx 1500$) in our program.

Many studies over the past decade have demonstrated that globular clusters are, in general, not comprised of single stellar populations as once supposed. Instead, any given cluster appears to possess multiple stellar generations with a variety of different properties (e.g., \citealt{Bedin:2004, Milone:2009, Gratton:2012, Milone:2012, Piotto:2015} for the Galactic clusters, \citealt{Dalessandro:2016,Niederhofer:2017a,Niederhofer:2017b} for the Small Magellanic Cloud cluster NGC 121). Future work that incorporates observations from our HST program using WFC3 and the F336W filter will examine the multiple population properties of our six target LMC clusters. However, the existence of multiple populations does not significantly affect cluster color-magnitude diagrams (CMDs) constructed using broadband optical filters such as F606W and F814W (except in extreme cases, such as $\omega$ Centauri, \citealt{Bedin:2004}), as the main effects of multiple populations are concentrated in the blue-uv part of the spectrum (e.g. \citealt{Sbordone:2011}). Thus, approximating both our LMC targets and the reference Galactic GCs as single populations for the present analysis is a valid approach.

This paper is arranged as follows. In Section 2, we discuss the dataset and the relevant photometry. We analyze the clusters to derive their relative ages in Section 3 and determine distances via subdwarf and HB fitting in Section 4. Section 5 lists our conclusions.


\section{Data}\label{Data}

\begin{figure}
  \centering
    \includegraphics[width=0.5\textwidth]{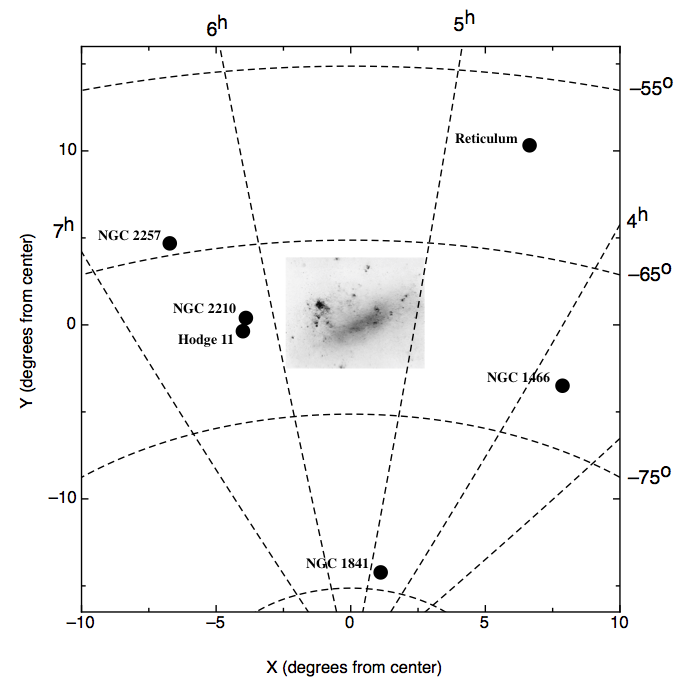}
  \caption{The LMC and surrounding region. The six clusters from HST Cycle 23 program GO-14164 (PI: Sarajedini) are indicated on the image.}
  \label{fig:lmc_locs}
\end{figure}

The data for our analysis come from HST Cycle 23 program GO-14164 (PI: Sarajedini). Deep imaging was obtained for six ancient LMC globular clusters -- NGC 1466, NGC 1841, NGC 2210, NGC 2257, Hodge 11, and Reticulum -- through the F606W and F814W filters with the Advanced Camera for Surveys (ACS) Wide Field Channel (WFC), and through the F336W filter with the Wide Field Camera 3 (WFC3) Ultraviolet and Visual (UVIS) channel. In this paper we will consider only the F606W and F814W data. These six clusters were chosen as they comprise the complete sample of LMC globular clusters outside the confusion-limited crowded stellar fields of the LMC bar region, as well as spanning a range of galactocentric radius, luminosity, and structure. The locations of the clusters relative to the LMC are shown in Figure \ref{fig:lmc_locs}.

Full details of the data acquisition and photometric analysis can be found in \cite{Mackey:2017}; here we provide a brief description for completeness. Each cluster was observed for two orbits in the F606W filter resulting in either 13 or 14 images (depending on the visibility of the target), and for three orbits in the F814W filter resulting in either 19 or 20 images. Of these image sets, two each per filter per cluster were short exposures ($\approx 50-70$s per frame) while the remainder were much longer ($\sim 350-520$s per frame). We used the {\sc dolphot} software package \cite[e.g.,][]{Dolphin:2000} to photometer the images, performing one run on the long exposures and one on the short exposures, and then merging the two quality-filtered output catalogues. To ensure that our photometry was minimally affected by imperfect charge transfer efficiency (CTE) in the ACS/WFC chips, we utilised the images from the {\sc calacs} pipeline corrected using the pixel-based algorithm of \citet{Anderson:2010}. Our final measurements are on the calibrated VEGAmag scale of \citet{Sirianni:2005}. For each cluster the photometry reaches from the top of the red-giant branch (RGB) to more than $4$ magnitudes below the main sequence turn off point (MSTOP). The signal-to-noise ratio around the HB level is $\sim 1900$, the MSTOP is typically $\sim 300$ per star, and remains at at least $\sim 30$ four magnitudes fainter than the MSTOP. 

For the LMC clusters analyzed herein, we adopt metallicities as summarized in Table \ref{tab:fehs}. The metallicity measurements are largely from high-resolution spectroscopic observations of these clusters, with the exception of NGC 1466. NGC 1466 has a photometric metallicity derived from RR Lyrae observations, and at present lacks spectroscopic measurements except for a handful of individual stars. Column 2 notes the metallicities from the cited references in column 3, which are provided in the metallicity scale noted in column 4. In order to adopt a consistent metallicity scale throughout this work, these [Fe/H] measurements are converted to the CG97 (\citealt{Carretta:1997}) metallicity scale, with the result provided in the final column of Table \ref{tab:fehs}; use of the CG97 metallicity scale provides consistency with the \cite{Marin-Franch:2009} study. The errors in the transformed metallicities include an assumed 0.2 dex measurement uncertainty as well as the propagated uncertainties of the transformation equations (\citealt{Carretta:1997, Carretta:2009}).

\begin{table*}
\caption{Assumed metallicities for our target LMC clusters}
\centering
\begin{threeparttable}[b]
    \begin{tabular}{@{}|l|c|l|l|c|@{}}
    \hline
\textbf{Cluster} & \textbf{[Fe/H]$_{ref}$} & \textbf{Reference} & \textbf{Original Scale} & \textbf{[Fe/H]$_{CG97}$} \\  
\hline
NGC 1466 	& -1.9*	& \protect\citet{Walker:1992a}	& ZW84	&  -1.70 $\pm$ 0.50	\\
NGC 1841 	& -2.02	& \protect\citet{Grocholski:2006}	& CG97	&  -2.02 $\pm$ 0.20	\\
NGC 2210 	& -1.65	& \protect\citet{Mucciarelli:2010}	& C09	&  -1.45 $\pm$ 0.20	\\
NGC 2257 	& -1.95	& \protect\citet{Mucciarelli:2010}	& C09	&  -1.71 $\pm$ 0.20	\\
Hodge11 	& -2.00	& \protect\citet{Mateluna:2012}	& C09	&  -1.76 $\pm$ 0.20	\\
Reticulum 	& -1.57	& \protect\citet{Grocholski:2006}	& CG97	&  -1.57 $\pm$ 0.20	\\
\hline
    \end{tabular}
    \begin{tablenotes}[b]
	\item $^*$ Measurement only given to one decimal precision by \protect\citet{Walker:1992a}. The propagation of error from the conversion of the measurement from the ZW84 scale to the CG97 scale leads to what a generous uncertainty in metallicity.
\end{tablenotes}
\end{threeparttable}
\label{tab:fehs}
\end{table*}


\section{Relative Ages}\label{Age}

To determine relative ages for the LMC clusters, we follow closely the procedure developed by \cite{Marin-Franch:2009} for their study of Galactic globular cluster relative ages. By measuring differences in the MSTOP apparent magnitude with respect to reference clusters, anchoring the apparent magnitudes to a distance scale, and comparing to expectations from theory, we determine precise relative ages for the LMC clusters. The details of our methodology are are outlined below. Throughout this process, we adopt the CG97 metallicity scale as discussed in Section \ref{Data}; the CG97 scale is also used in \cite{Marin-Franch:2009}.

A mean ridge line (MRL) is derived to represent the fiducial line for each cluster. As in \cite{Marin-Franch:2009}, we determine mean ridge lines for the F606W$-$F814W, F814W CMD because the sub-giant branch is more vertical and the results are improved with respect to using F606W as the magnitude. However, in the rest of the analysis, the F606W$-$F814W, F606W CMD is used to derive the MSTOP in F606W magnitude (\citealt{Marin-Franch:2009}). To determine the MRL, we use a moving bin in magnitude with a size of 0.5 mag and steps of 0.02 magnitude. The MRL location is determined from the average of each bin, rejecting 3-$\sigma$ outliers to determine a MRL. We repeat this MRL-derivation process, shifting the bin locations each time, to lessen dependence of the result on the bin centers. Over five iterations, the derived MRLs are combined; a smoothed radial basis function is used to represent the final fiducial line.

The magnitude of the MSTOP is found by taking the bluest point of a spline fit locally to the MRL in the MSTOP region. The MSTOP location is re-determined ten times, each time offsetting the MRL bins by 0.01 mag; the adopted MSTOP magnitude for the cluster is calculated as the mean of these ten determinations of the bluest point of the spline. Through this process, we determine apparent magnitudes in F606W of the MSTOP for the six LMC clusters, provided in the fourth column of Table \ref{tab:rel_ages}. 

Because this is a relative analysis, the MSTOPs of the LMC clusters need to be compared to reference MSTOP locations. Two particular regions along the fiducial lines were designated by \cite{Marin-Franch:2009} to be minimally affected by age; specifically, on the main sequence (M$_{\text{MSTOP, F606W}}$+3$\leq$ m $\leq$M$_{\text{MSTOP, F606W}}$+1.5) and on the red giant branch (M$_{\text{MSTOP, F606W}}$-2.5 $\leq$ m $\leq$M$_{\text{MSTOP, F606W}}$-1.5). Because these RGB and MS regions are largely unaffected by age, a comparison of MSTOP magnitude between the reference and MRL gives an indication of the relative age of two clusters.  The MRL of these two locations are used to ``shift" the entire MRL of a reference cluster in color and magnitude to match the same regions in each LMC cluster. In doing so, any differences in line of sight reddening between the clusters are automatically accounted for, making the age determinations free of assumptions of distance or reddening. By using the GGC clusters as reference clusters, we thus determine the differences in MSTOP magnitudes with respect to the LMC clusters.

\begin{figure*}
  \centering
    \includegraphics[width=\textwidth]{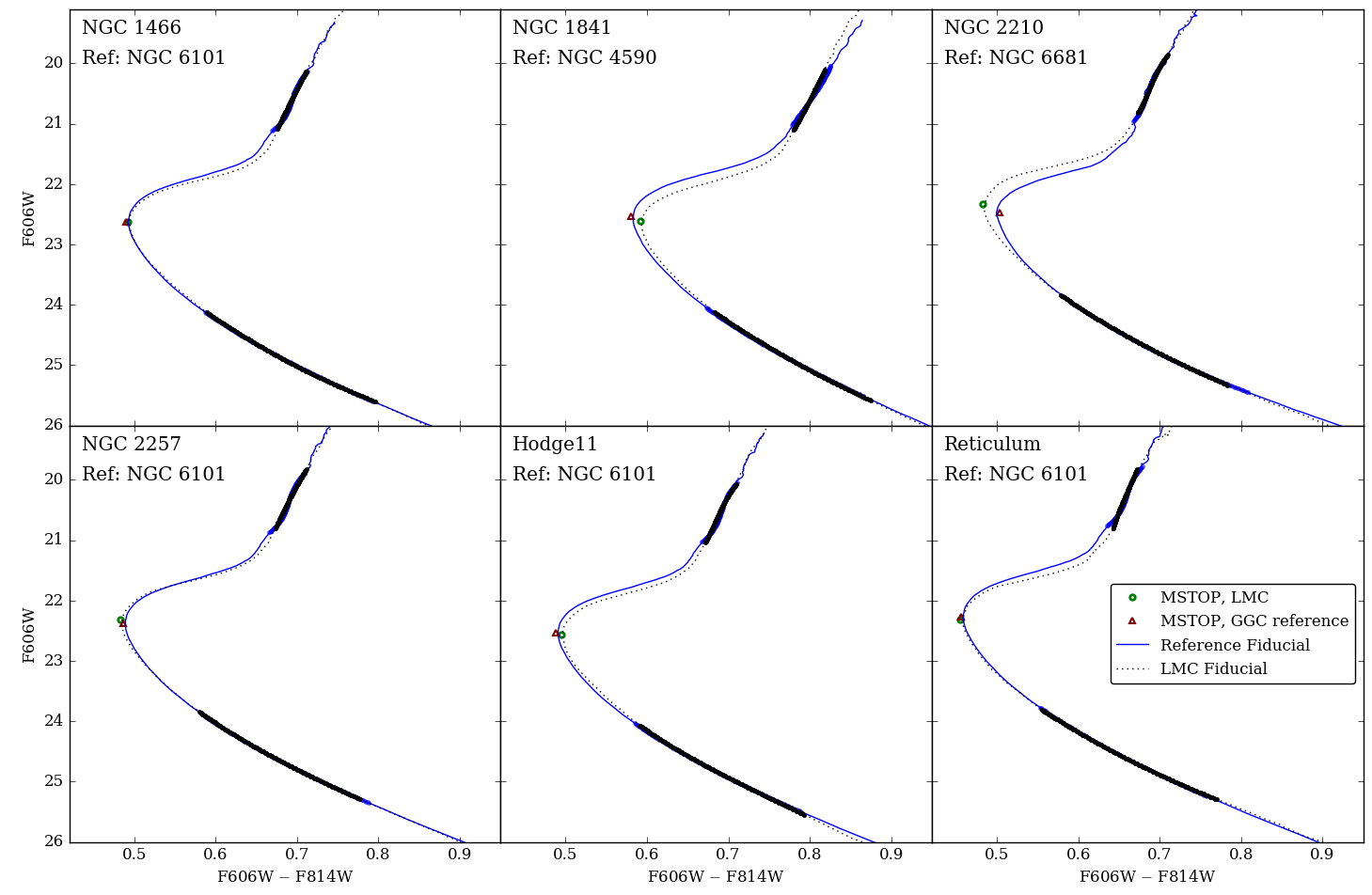}
  \caption{A comparison of the LMC fiducial sequences (dotted black lines) to the shifted reference Galactic cluster fiducials (solid blue lines, see Table \ref{tab:rel_ages}). The intervals upon which the least-squares match is performed are indicated by the thick regions for each fiducial. The location of the LMC MSTOP is indicated by the open green circle and by the maroon triangle for the Galactic reference cluster MSTOP location.}
  \label{fig:rel_ages_shifts}
\end{figure*}

We use the same GGC reference clusters as Marin-Franch et al. (2009) across the following metallicity ranges: NGC 6981 (--1.3 $\leq$ [Fe/H] $\textless$ --1.1), NGC 6681 (--1.5 $\leq$ [Fe/H] $\textless$ --1.3), NGC 6101 (--1.8 $\leq$ [Fe/H] $\textless$ --1.5), and NGC 4590 (--2.3 $\leq$ [Fe/H] $\textless$ --1.8). We use a least squares approach to minimize offsets between the MRL of the reference clusters and the LMC clusters in the intervals defined by \cite{Marin-Franch:2009} and described above. The result of this process is demonstrated in Figure \ref{fig:rel_ages_shifts}, where each LMC cluster (black dotted lines) is compared to the relevant GGC reference (blue solid lines). The fiducials of the GGC reference clusters are shifted to match the LMC clusters in the regions of the CMD discussed above; these regions are also highlighted in Figure \ref{fig:rel_ages_shifts}. The MSTOP for the LMC (green) and reference Galactic cluster (maroon) are also indicated in the plot. While the fiducials for the LMC clusters are largely consistent with their GGC reference cluster counterpart, we note that there is a slight mismatch between the NGC 1841 and NGC 4590 fiducials on the upper red giant branch. This does not appear to be driven by our approach but may be due to differing metallicities or [$\alpha$/Fe] abundances between the two clusters.

As in \cite{Marin-Franch:2009}, shifts among the set of GGC reference clusters are determined by matching MRL for the reference clusters in adjacent metallicity ranges. The absolute MSTOP magnitude in F606W for NGC 6752 of 3.87 $\pm$ 0.15 for calibration is adopted, as are comparable uncertainties arising from the methodology, to be consistent with \cite{Marin-Franch:2009}. This allows us to obtain the absolute magnitude of the MSTOP, given in the fifth column of Table \ref{tab:rel_ages}. As the uncertainty in the distance of NGC 6752 affects the absolute age scale rather than the relative age scale, we do not include this in our error budget. In re-deriving cluster ages from \cite{Marin-Franch:2009} for clusters with [Fe/H] $\leq$ --1.5, we find a minimal mean offset of the MSTOP magnitudes of  0.01 $\pm$ 0.03; this is incorporated into our cited errors.

\renewcommand{\arraystretch}{1.25}
\begin{table*}
\caption{Relative ages for our LMC cluster sample}
\centering
\begin{threeparttable}[b]
    \begin{tabular}{@{}|l|c|c|c|c|c|c|@{}}
    \hline
\textbf{Cluster} & \textbf{[Fe/H]$_{CG97}$$^a$} & \textbf{Ref. Cluster$^b$} & \textbf{m$_{\text{F606W, MSTOP}}$} & \textbf{M$_{\text{F606W, MSTOP}}$} & \textbf{Relative Age (Gyr)} & \textbf{Absolute Age (Gyr)}    \\  
\hline
NGC 1466 	&  -1.70 	&  NGC 6101 	& 22.62 $\pm$ 0.02  &  3.75 $\pm$ 0.18 		&  12.18 $^{+ 0.95 }_{ -1.12 }$    	&  13.38 $^{+ 1.67 }_{ -2.28 }$	\\
NGC 1841 	&  -2.02 	&  NGC 4590 	&  22.61 $\pm$ 0.02 &  3.64 $\pm$ 0.17		&  12.57 $^{+ 0.47 }_{ -1.57 }$    	&  13.77 $^{+ 1.05 }_{ -2.41 }$	\\
NGC 2210 	&  -1.45 	&  NGC 6681 	&  22.34 $\pm$ 0.02 &  3.70 $\pm$ 0.18 		&  10.43 $^{+ 1.15 }_{ -1.06 }$    	&  11.63 $^{+ 1.80 }_{ -1.12 }$	\\
NGC 2257 	&  -1.71 	&  NGC 6101 	&  22.32 $\pm$ 0.01 &  3.69 $\pm$ 0.18 	 	&  11.54 $^{+ 1.08 }_{ -1.28 }$    	&  12.74 $^{+ 1.87 }_{ -2.18 }$	\\
Hodge11 	&  -1.76 	&  NGC 6101 	&  22.56 $\pm$ 0.01 &  3.78 $\pm$ 0.17 		&  12.72 $^{+ 0.65 }_{ -0.85 }$    	&  13.92 $^{+ 1.48 }_{ -2.01 }$	\\
Reticulum 	&  -1.57 	&  NGC 6101 	&  22.31 $\pm$ 0.01 &  3.79 $\pm$ 0.17 		&  11.89 $^{+ 0.87 }_{ -0.82 }$    	&  13.09 $^{+ 2.21 }_{ -1.98 }$	\\
\hline
    \end{tabular}
    \begin{tablenotes}[b]
	\item $^a$ From Table \ref{tab:fehs}.
	\item $^b$ As in \protect\citet{Marin-Franch:2009}.
\end{tablenotes}
\end{threeparttable}
\label{tab:rel_ages}
\end{table*}
\renewcommand{\arraystretch}{1.1}

\begin{figure}
  \centering
    \includegraphics[width=0.5\textwidth]{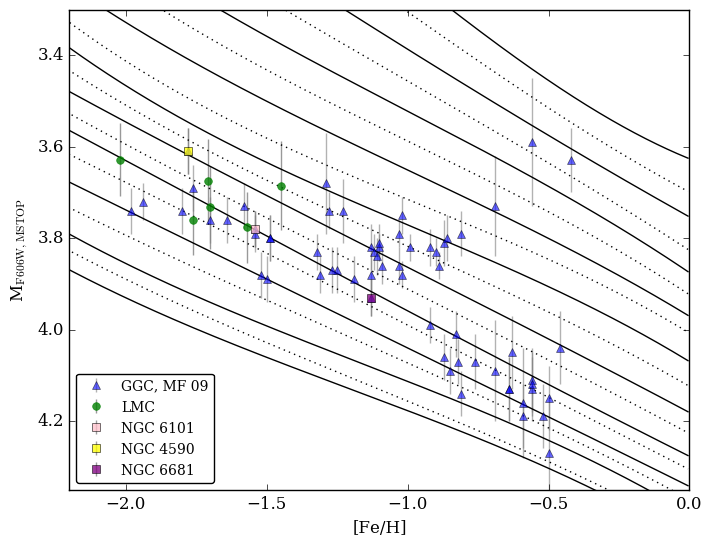}
  \caption{The magnitude of the MSTOP of the LMC clusters (green circles) according to their assumed metallicities. The MSTOP determined for Galactic globular clusters from \protect\cite{Marin-Franch:2009} are plotted as blue triangles. The solid lines delineate the MSTOP from the theoretical models (\protect\citealt{Dotter:2008}) from 6 Gyr (upper right) to 15 Gyr (lower left) at 0.5 Gyr intervals, alternating solid and dashed lines for clarity. The Galactic reference clusters are indicated by the colored squares. The theoretical grid assumes [$\alpha$/Fe] = 0.2.}
  \label{fig:rel_ages}
\end{figure}

We plot the MSTOP magnitude estimates for the six LMC clusters alongside the GGC MSTOP magnitudes from \cite{Marin-Franch:2009} in Figure \ref{fig:rel_ages}. From a grid of DSED isochrones at 0.1 dex metallicity intervals, the smoothed, diagonal lines in this figure indicate MSTOP locations for isochrones ranging in age from 6 Gyr (upper right) to 15 Gyr (lower left) in steps of 0.5 Gyr.  For clarity, we alternate between solid and dashed lines. Additionally, the LMC clusters show enhancement in [$\alpha$/Fe] comparable to the enhancement seen in the Galactic globular clusters (\citealt{Mucciarelli:2010}). The grid in Figure \ref{fig:rel_ages} is generated assuming an $\alpha$-enhanced model ([$\alpha$/Fe] = 0.2). The grid of \cite{Dotter:2008} DSED isochrones are used to derive MSTOP in the same iterative process as described above - fitting a spline locally to the MSTOP and determining the bluest point. This facilitates a comparison between the calibrated MSTOP magnitudes from the observed clusters to the theoretical models to determine ages. We interpolate between the models to determine ages for the six LMC clusters, presented in Table \ref{tab:rel_ages}.

The relative ages we derive are consistent with the LMC clusters having similar properties to the old, inner halo Milky Way clusters. However, we note that the metallicities of the LMC clusters are not fully agreed upon in the literature; if our adopted metallicities are inaccurate then our conclusions on the relative ages of these clusters may need re-examination. We estimate a change of ~0.1 dex in metallicity leads to a change of up to $\approx$ 0.4 Gyr in relative age. Nevertheless, our relative age determinations are a significant improvement on more uncertain age estimates from previous studies of these clusters (\citealt{Brocato:1996, Olsen:1998, Piatti:2009, Jeon:2014}). Through our analysis, relative ages are determined with precision between 6\% (Hodge 11) and 10\% (NGC 2210), with a mean precision across the sample of 8.4\%.

With our substantial improvement in LMC cluster age measurements over previous studies, we can re-examine the coevality of globular cluster formation in the MWG and the LMC - two galaxies that are thought to have been widely separated at the epoch of cluster formation (\citealt{Kallivayalil:2006,Kallivayalil:2013}), and which thus present independent and very different formation environments, as the MWG is a more massive and metal-rich galaxy than the LMC. Previous work has indicated that early cluster formation in the Milky Way and LMC was broadly concurrent within 1.5 to 2 Gyr (\citealt{Johnson:1999,Olsen:1998,Brocato:1996,Mackey:2004b}).

We propagate the individual errors of the relative age estimates from Table \ref{tab:rel_ages} and find an average age for the LMC clusters of 11.8 $\pm$ 0.4 Gyr (0.8 Gyr stddev). For the Galactic clusters in a similar metallicity range ([Fe/H] $\leq$ --1.5), the mean age is 12.0 $\pm$ 0.2 Gyr (0.6 Gyr stddev). Consequently, this suggests the two sets of metal-poor old clusters are coeval within 0.2 $\pm$ 0.4 Gyr (1.0 Gyr stddev). As the best test of the synchronicity of globular cluster formation in two completely unrelated galaxies so far, we find clear evidence that the first epoch of GC formation occurred essentially simultaneously in these two completely different galactic environments.

The metal-poor Sagittarius clusters have been found to also be very similar to those in our Galaxy (\citealt{Mackey:2004,Marin-Franch:2009}). Terzan 8 is the one cluster in the Sagittarius dwarf galaxy that is more metal-poor than -1.5 on the CG97 scale. \cite{Marin-Franch:2009} find the age of this cluster to be within 0.5 +/- 0.5 Gyr of the mean value we quote above. However, this trend does not necessarily seem to be consistent in the Small Magellanic Cloud (SMC). While NGC 121 is the only ``old" cluster in the SMC, the majority of recent work suggests it is younger than the oldest Galactic clusters despite having similar metallicity; estimates put the age in the range of 10.5 to 11.8 Gyr, depending on study and model choice (\citealt{Glatt:2008, Dolphin:2001, Mighell:1998}). However, NGC 121 has not been measured on the same relative age scale we use herein and a direct comparison is incomplete in this respect. Although the MWG, LMC, and Sagittarius appear to have closely synchronous initial metal-poor GC formation, the example of the SMC indicates that this may not be universal.

The ages derived here are relative ages, not absolute ages. To compare ages directly with previous studies, a reference absolute age must be adopted. From the absolute ages of \cite{OMalley:2017}, we use the mean absolute age of the GGC reference clusters to calibrate our LMC relative ages to absolute ages. Specifically, \cite{OMalley:2017} found ages of 12.4$\pm$1.3 Gyr for NGC 4590, 13.4$\pm$1.5 Gyr for NGC 6101, and 12.5$\pm$1.7 Gyr for NGC 6681 and we adopt a mean age of 12.8$\pm$0.9 Gyr (note that this is the same absolute age adopted by \citealt{Marin-Franch:2009} as well). This absolute age is compared to the mean relative ages of these same clusters (11.6 Gyr); the LMC clusters' relative ages are adjusted according to the difference of 1.2 Gyr. These values are listed in the final column of Table \ref{tab:rel_ages} and include the uncertainty of 0.15 mag in the absolute magnitude of the MSTOP for NGC 6752, the zeropoint cluster. 

\begin{figure*}
  \centering
    \includegraphics[width=\textwidth]{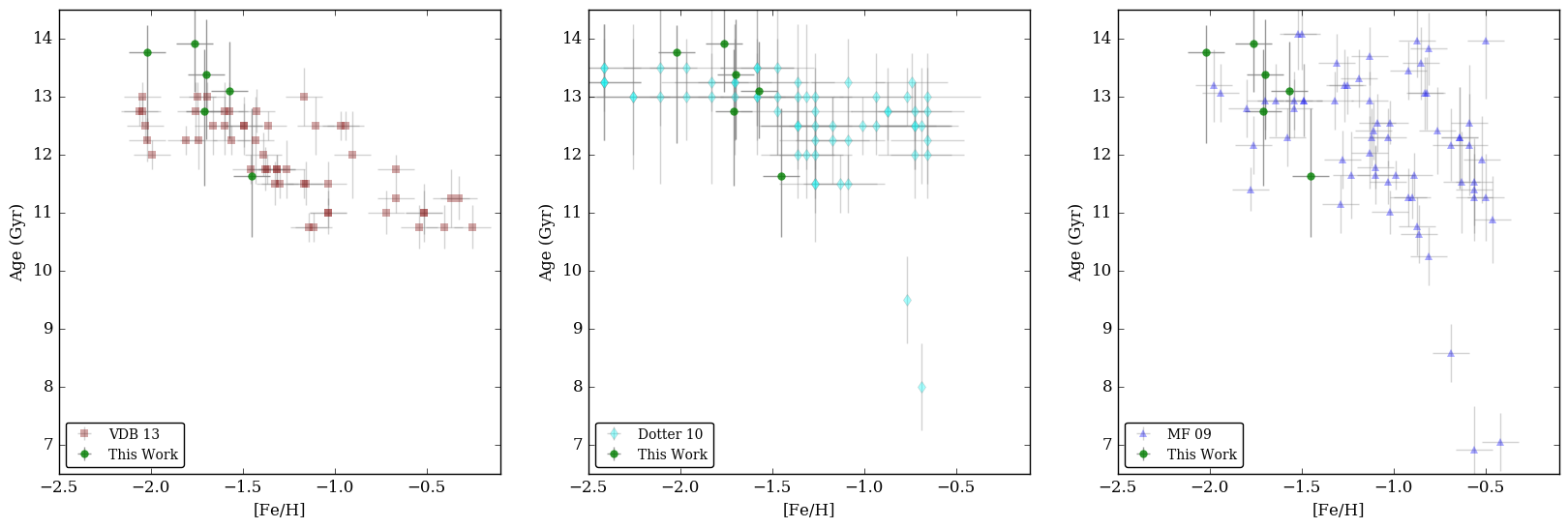}
  \caption{The age-metallicity relation for the Galactic globular clusters and the LMC clusters, compared. The leftmost panel shows ages from \protect\cite{Vandenberg:2013} (red squares), the middle panel shows ages from \protect\cite{Dotter:2010} (cyan diamonds), and the ages from \protect\cite{Marin-Franch:2009} (blue triangles). Our results may be compared directly to \protect\cite{Marin-Franch:2009}, and qualitatively with \protect\cite{Vandenberg:2013} and \protect\cite{Dotter:2010}, who employ different methods and age calibrations.}
  \label{fig:amr}
\end{figure*}

Using these absolute ages, the age-metallicity relation (AMR) is reconstructed in Figure \ref{fig:amr} using ages of the Galactic clusters from \cite{Vandenberg:2013} (left), \cite{Dotter:2010} (middle), and \cite{Marin-Franch:2009} (right, assuming reference age of 12.8 Gyr). The LMC clusters are plotted for comparison (green circles). Our results may be compared directly to \cite{Marin-Franch:2009}. With respect to \cite{Vandenberg:2013} and \cite{Dotter:2010}, the comparisons are qualitative, as different methods and age calibrations may lead to vertical offsets in Figure \ref{fig:amr}.

As in Figure \ref{fig:rel_ages}, the LMC clusters appear to more closely mimic the old, metal-poor inner halo Galactic clusters. NGC 2210 could plausibly be younger than the other LMC clusters, although this is not a statistically significant deviation. Nonetheless, from inspection of the AMR, it is possible that NGC 2210 may be more similar to Galactic clusters belonging to the younger branch of the AMR (the outer halo; presumably accreted clusters) than the older branch (the inner halo; likely \emph{in-situ} clusters). Although there is not sufficient evidence here to make a strong conclusion, it is certainly of interest to pursue in future work in order to better understand the formation history of the LMC and its globular cluster system.

\section{Distances}\label{Distances}


\subsection{Horizontal Branch}\label{sec:zahb}

\begin{figure*}
  \centering
    \includegraphics[width=\textwidth]{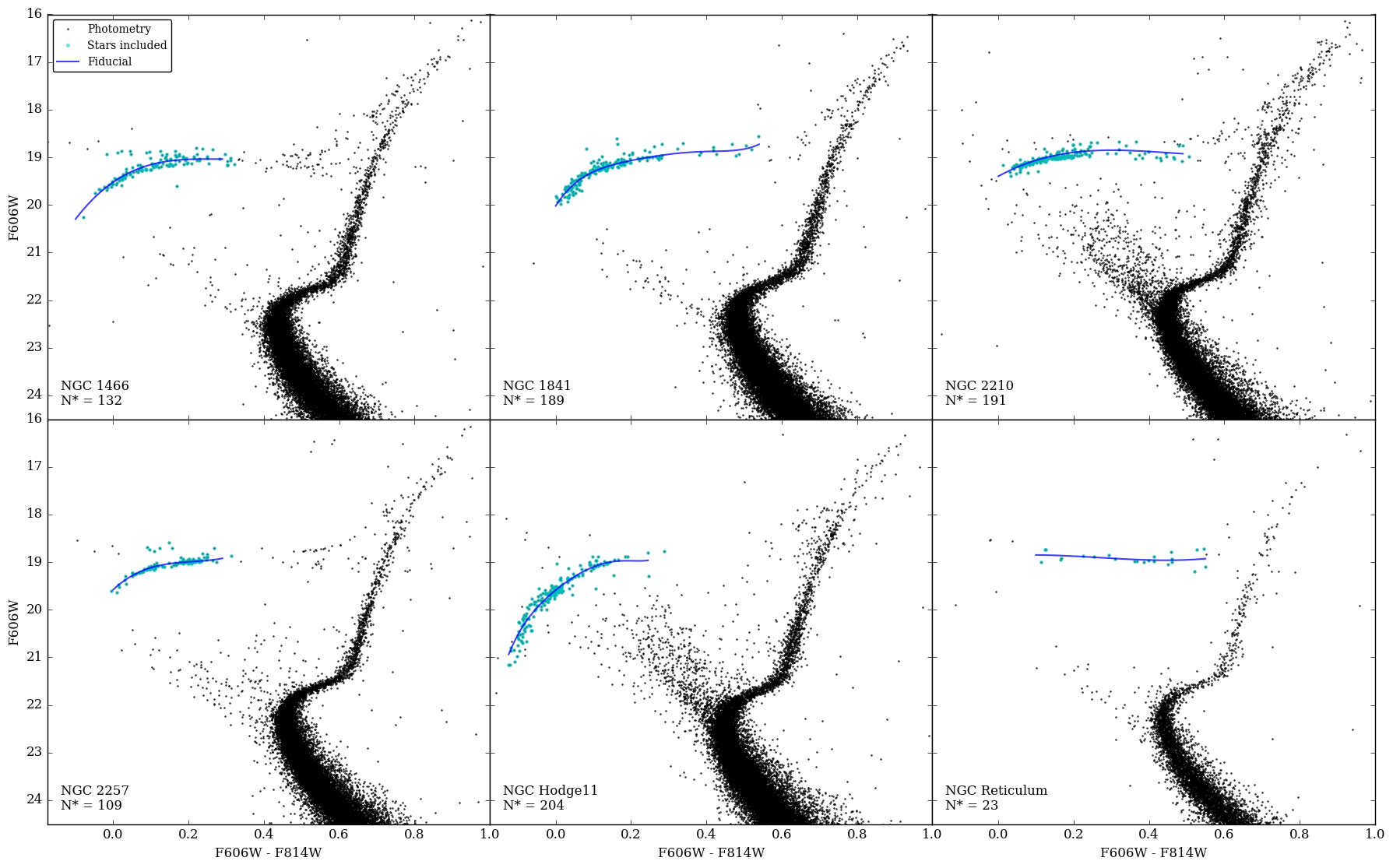}
  \caption{Fiducial horizontal branches for the six LMC clusters in our sample. In each panel, the HST photometry is plotted in black, the included stars are the larger cyan points, and the fiducial spline (see text for details) is indicated by the solid blue line. The cluster name and number of stars (N*) included in the fit are indicated in the lower left of each panel.}
  \label{fig:zahbLMC}
\end{figure*}

To obtain distance estimates from observations of the horizontal branch (HB), fiducials of the HBs are created for the LMC clusters and compared to several reference GGC clusters. We initially estimate the cluster HB fiducial by eye and fit a cubic spline to the points. To iterate on this and obtain a better estimate, the HB is binned in intervals of color of 0.04 along the initial spline, overlapping by 0.01. In each color bin, using stars within a 3-$\sigma$ range of the initial spline, we determine the average magnitude of the HB in that bin. These results are used to redetermine the HB with a cubic spline, as seen by the solid blue line in Figure \ref{fig:zahbLMC} for each LMC cluster, where the included HB stars are indicated (cyan points). We note that while blue straggler stars or field stars may contaminate the included HB stars, their low frequency is unlikely to skew the estimated HB fiducials.

This process is repeated for six GGC reference clusters to serve as comparisons to the HB fiducials of the LMC clusters. These Galactic clusters are chosen to (i) have low reddening, (ii) have a populated HB, (iii) have available photometry in the same HST filters from \cite{Sarajedini:2007}. Using the same method described above, we derive HB fiducials for two moderately metal-rich Galactic clusters (NGC 5272, NGC 6584), two intermediate metallicity clusters (NGC 5024, NGC 6809), and two metal-poor clusters (NGC 4590, NGC 6341). We adopt the foreground reddenings and distance moduli from the \cite{Harris:2010} catalogue\footnote{\url{http://physwww.mcmaster.ca/~harris/mwgc.dat}} for these reference clusters; these values are listed in Table \ref{tab:GGCref}. We note that \cite{Harris:2010} derive their distances from the assumption that M$_V$(HB) = 0.16 [Fe/H] + 0.84 and their E(B--V) from an average of several measurements; we assume a reddening law of R$_V$=3.1 (\citealt{Cardelli:1989}). Additionally, the \cite{Harris:2010} catalogue metallicities are on the C09 scale; for consistency (as in Section \ref{Data}) we convert these to the CG97 scale, and these values are listed in Table \ref{tab:GGCref}.

\begin{figure*}
  \centering
    \includegraphics[width=0.65\textwidth]{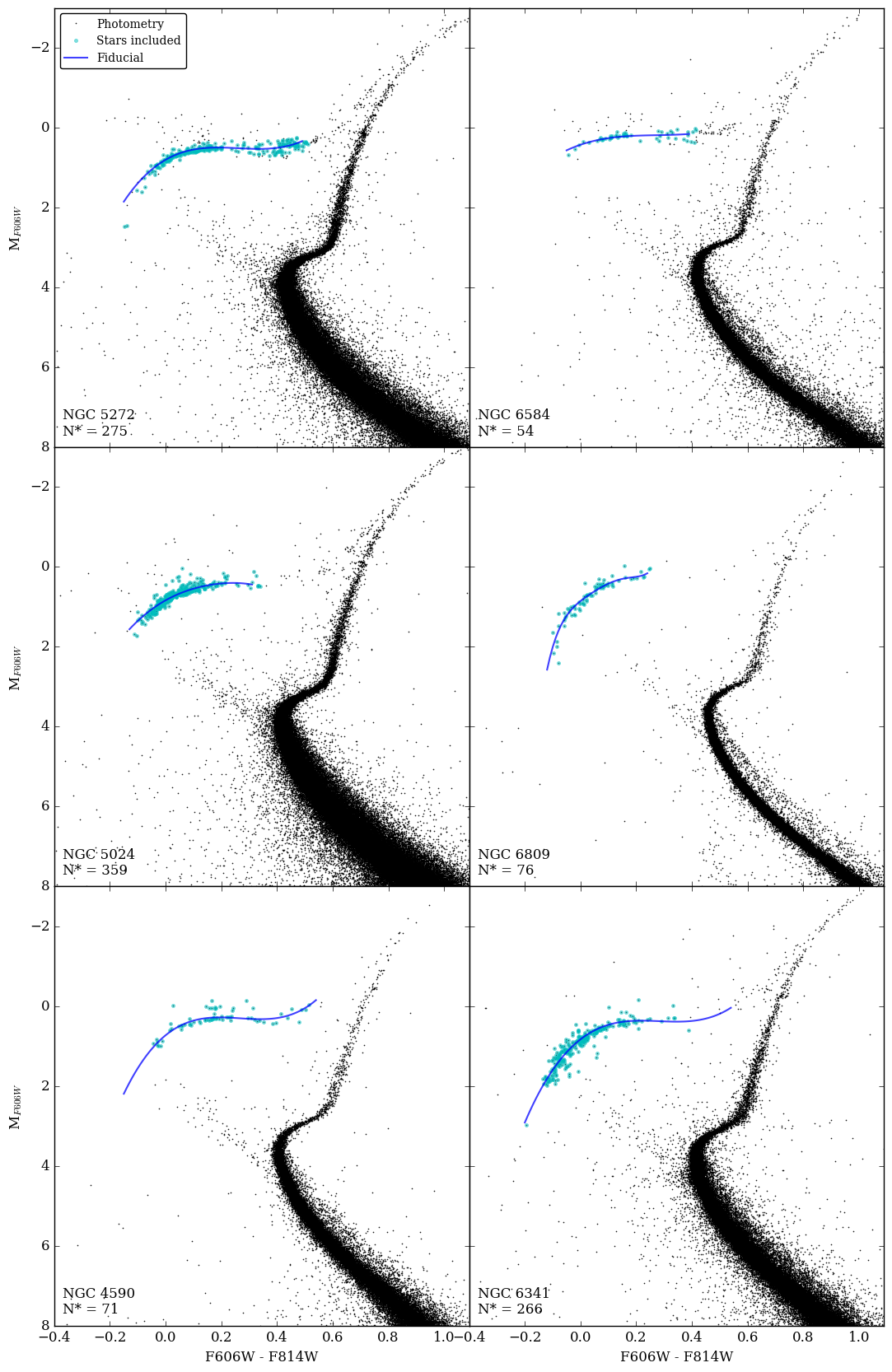}
  \caption{Fiducial horizontal branches for the reference Galactic globular clusters in our sample. In each panel, the HST photometry is plotted in black, the included stars are the larger cyan points, and the fiducial spline (see text for details) is indicated by the solid blue line. The cluster name and number of stars (N*) included in the fit are indicated in the lower left of each panel.}
  \label{fig:zahbGGC}
\end{figure*}

\begin{table}
\caption{GGC HB Calibration Clusters}
\centering
    \begin{tabular}{@{}|l|c|c|c|@{}}
    \hline
\textbf{Cluster} & \textbf{[Fe/H]$_{CG97}$} & \textbf{E(B--V)} & \textbf{Distance Modulus}   \\  
\hline
NGC 5272	& -1.32 $\pm$ 0.19	& 0.01	& 15.07		\\
NGC 6584	& -1.32 $\pm$ 0.20	& 0.10	& 15.96		\\
NGC 5024	& -1.84 $\pm$ 0.20	& 0.02	& 16.32		\\
NGC 6809	& -1.70 $\pm$ 0.20	& 0.08	& 13.89		\\
NGC 4590	& -1.96 $\pm$ 0.20	& 0.05	& 15.21	 	\\
NGC 6341	& -2.03 $\pm$ 0.21	& 0.02	& 14.65		\\
        \hline
    \label{tab:GGCref}
    \end{tabular}
\end{table}

In order to obtain distance measurements, we assume reddenings for the LMC clusters from Walker (1992, 1993), as indicated in Table \ref{tab:zahb_dists}. While an assumption of reddening is necessary to derive distances, the accuracy of the adopted reddening values could plausibly affect our distance estimates. However, the reddening estimates we adopt are found to differ on average from the foreground reddening estimates of \cite{Schlegel:1998} and \cite{Schlafly:2011} by $\sim$ 0.006.

With the derived HB fiducials, the fiducials from the GGC reference clusters are shifted to match those of the LMC clusters in a least-squares fashion. The resulting shift in magnitude from this fit, in conjunction with the assumed GGC distances from Table \ref{tab:GGCref}, determines the distance for each LMC cluster. Each LMC cluster has its distance derived with respect to two reference GGC cluster fiducials, as indicated in the fourth column of Table \ref{tab:zahb_dists}. The final column of this table provides the average of those two distances.

The cited uncertainties on the derived distances in Table \ref{tab:zahb_dists} incorporate the scatter (standard deviation in magnitude) and number of stars used to derive each LMC fiducial, the scatter and number of stars for the GGC fiducials, and the errors from the least-squares fitting between the GGC and LMC fiducials.

\begin{table*}
\caption{Distance Estimates: ZAHB}
\centering
\begin{threeparttable}[b]
    \begin{tabular}{@{}|l|c|c|c|c|c|@{}}
    \hline
\textbf{Cluster} & \textbf{[Fe/H]$_{CG97}$ $^a$}	  & \textbf{Assumed E(B--V)$^b$} & \textbf{E(B--V)$^c$} & \textbf{Reference HBs} & \textbf{$\mu_{ZAHB}$}  \\  
\hline
NGC 1466 & -1.70	& 0.09	& 0.07	& NGC 5024, NGC6809	& 18.67 $\pm$ 0.08	\\
NGC 1841 & -2.02	& 0.18	& 0.19	& NGC 4590, NGC6341	& 18.58 $\pm$ 0.05	\\
NGC 2210 & -1.45	& 0.06	& 0.08	& NGC 5272, NGC6584	& 18.52 $\pm$ 0.04	\\
NGC 2257 & -1.71	& 0.04	& 0.06	& NGC 5024, NGC6809	& 18.61 $\pm$ 0.07	\\
Hodge 11   & -1.76	& 0.08	& 0.08	& NGC 5024, NGC6809	& 18.57 $\pm$ 0.08	\\
Reticulum & -1.57	& 0.03	& 0.02	& NGC 5272, NGC6584	& 18.57 $\pm$ 0.06	\\	
        \hline
    \label{tab:zahb_dists}
    \end{tabular}
\begin{tablenotes}[b]
	\item $^a$ As in Table \ref{tab:fehs}.
	\item $^b$ Absorptions adopted from Walker (1992, 1993), assuming R$_{V}$ =3.1.
	\item $^c$ Absorptions from \cite{Schlegel:1998} for comparison, assuming R$_{V}$ =3.1.
\end{tablenotes}
\end{threeparttable}
\end{table*}

\begin{figure*}
  \centering
    \includegraphics[width=\textwidth]{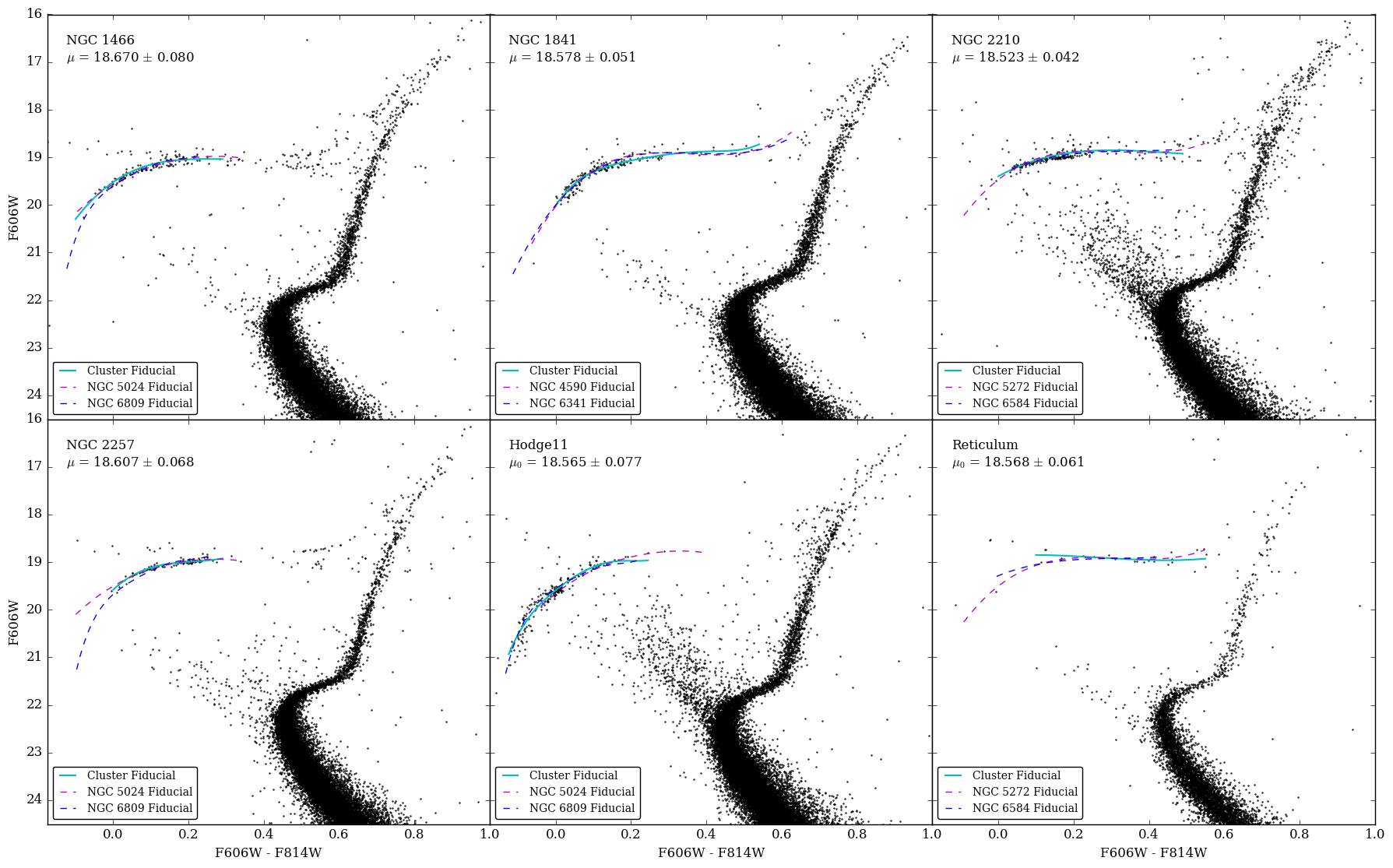}
  \caption{In each panel, the HST photometry is plotted in black. The shifted HB fiducials of the reference clusters are indicated by the dashed blue and magenta lines. The solid cyan line indicates the fiducial of the LMC cluster.}
  \label{fig:zahb}
\end{figure*}

Through this process, we obtain an average distance of \LMCdistHB across the six clusters, in line with previous studies for the LMC distance of approximately $\mu$ = 18.49 $\pm$ 0.09 (\citealt{de-Grijs:2014}). Further distance determination comparisons are discussed at the end of Section \ref{sec:subdwarfs}.


\subsection{Subdwarf Analysis}\label{sec:subdwarfs}

As the most well-understood phase of stellar evolution, fitting of main-sequence stars offers an opportunity to provide excellent distance measurements. Using parallaxes and highly precise HST photometry of local subdwarfs from GO-11704 (\citealt{Chaboyer:2017}), we leverage the accurately measured subdwarf absolute magnitudes and the clean main sequences of the six LMC clusters to determine distances. Details of the four subdwarfs in the metallicity range of the LMC clusters are presented in Table \ref{tab:subdwarf_details}. The magnitudes and colors of these stars are adjusted for their individual reddening, noted in Column 5.

In order to compare the observed main sequences of the LMC clusters to the photometry for the local subwarfs, it is necessary to adjust the colors of the latter to account for the various metallicities of the clusters and the subdwarfs. First, the metallicities of the individual stars are adjusted for their unique [$\alpha$/Fe] abundances using the following relation from \cite{Salaris:1993} and \cite{Dotter:2010}.

\begin{equation}\label{eq:alpha}
[M/H] = [Fe/H]+\log(0.638 \times 10^{[\alpha/Fe]} + 0.362)
\end{equation}

A grid of DSED isochrones for the HST/ACS filters over a metallicity range of -2.5 to -0.5 and an absolute magnitude range of 4.5 $\textless$ M$_{F606W}$ $\textless$ 7 is used. In this grid, we interpolate based on the metallicity of each star to determine the change in color necessary to correct the F606W$-$F814W color of the subdwarf to the reference metallicity of each of the clusters. The reference metallicity of each LMC cluster is also corrected as in Equation \ref{eq:alpha} for an assumed $\alpha$-enhancement of [$\alpha$/Fe] = 0.2.

\begin{figure*}
  \centering
    \includegraphics[width=\textwidth]{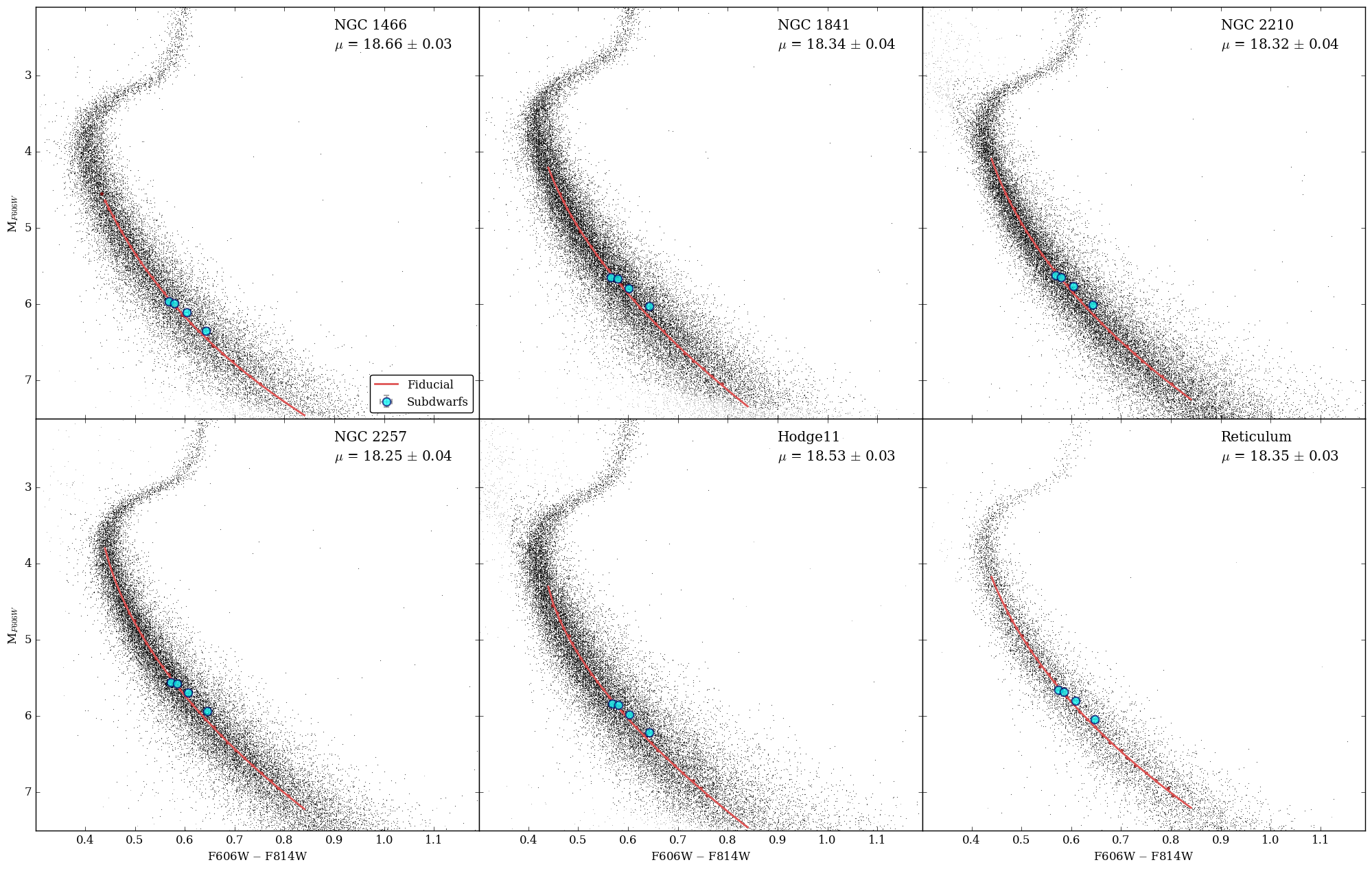}
  \caption{Subdwarf fitting to the main sequences of the LMC clusters. The photometry is shown in black. A fiducial, in red, is fit to the de-reddened cluster photometry. The subdwarfs, shown in cyan, are offset in magnitude to minimize the standard deviation to the fiducial, resulting in an estimate of the cluster distance.}
  \label{fig:subdwarfs}
\end{figure*}

Assuming a reddening for each LMC cluster (\citealt{Walker:1992, Walker:1993}, see Table \ref{tab:subdwarfs_res}), a main sequence fiducial is fit with a power law for cluster stars between an absolute F606W magnitude of 4.5 and 7 in overlapping magnitude bins of 0.1 mag. The fiducials are shown in the panels of Figure \ref{fig:subdwarfs} as red curves for each cluster. The fiducial is shifted via least squares to minimize the offsets of the HST parallax stars from the fiducial line in magnitude. The resulting distance moduli are given in the final column of Table \ref{tab:subdwarfs_res}, with the quoted error representing the uncertainty in the fiducial fit.

We find an average distance modulus to the LMC of \LMCdistSD (\LMCdistSDKpc). This is shorter than the mean distance we determined from the HB analysis in section \ref{sec:zahb} of \LMCdistHB. The two estimates are within 3-$\sigma$ but discrepant. This is likely due to different underlying assumptions, as discussed further at the end of this section. From the HB and subdwarf distances, we derive an average distance to the LMC of \LMCavgdist  (\LMCavgdistKpc).

\begin{table*}
\caption{Properties of the Reference Subdwarfs used for Cluster Distance Measurements}
\centering
    \begin{tabular}{@{}|l|c|c|c|c|c|c|@{}}
    \hline
\textbf{ID} & \textbf{$\Pi$ (mas)}  & \textbf{F606W} & \textbf{F814W}  & \textbf{E(B-V)} & \textbf{[Fe/H]}  & \textbf{[$\alpha$/Fe]}   \\  
\hline
HIP46120   &  14.49 $\pm$ 0.16     & 9.938 $\pm$ 0.0015     & 9.371 $\pm$ 0.0017      & 0.00	& -2.24    &  0.29 \\
HIP103269 &  14.12 $\pm$ 0.10    & 10.084 $\pm$ 0.0034    & 9.503 $\pm$ 0.0027   & 0.00	& -1.85   & 0.06 \\ 
HIP106924 &  14.48 $\pm$ 0.12    & 10.156 $\pm$ 0.0024     & 9.555 $\pm$ 0.0048   & 0.00	& -2.22   & 0.23 \\ 
HIP108200 &  12.39 $\pm$ 0.08   & 10.785 $\pm$ 0.0031    & 10.143 $\pm$ 0.0055   & 0.02	& -1.83    &0.01 \\ 
        \hline
    \label{tab:subdwarf_details}
    \end{tabular}
\end{table*}

\begin{table*}
\caption{Distance Estimates: Subdwarf Fitting}
\centering
\begin{threeparttable}[b]
    \begin{tabular}{@{}|l|c|c|c|c|c|c|@{}}
    \hline
\textbf{Cluster} & \multicolumn{2}{c}{\textbf{Assumed Values}} & \textbf{}  & \textbf{} & \textbf{}  & \textbf{} \\ \cline{2-3}
\textbf{Name} & \textbf{E(B-V)$^a$} &  \textbf{[Fe/H]$^b$}  & \textbf{$\mu$}  & \textbf{D (kpc)} & \textbf{$\mu_{0}$}  & \textbf{D$_0$ (kpc)}   \\  
\hline
NGC 1466 & 0.09   &  -1.7		&  18.66 $\pm$ 0.03 & 53.94 $\pm$ 0.80	& 18.40 $\pm$ 0.06 	& 47.90 $\pm$ 1.29 \\
NGC 1841 & 0.18  &  -2.02    		& 18.34 $\pm$ 0.04  & 46.61 $\pm$ 0.86	& 17.85 $\pm$ 0.11	& 37.10 $\pm$ 1.89   \\
NGC 2210 & 0.06    &  -1.45		& 18.32 $\pm$ 0.04  & 46.05 $\pm$ 0.80   	& 18.34 $\pm$ 0.02	& 46.58 $\pm$ 0.38	\\
NGC 2257 & 0.04  &  -1.71    		& 18.25 $\pm$ 0.04  & 44.60 $\pm$ 0.82	& 18.39 $\pm$ 0.04 	& 47.59 $\pm$ 0.86   \\
Hodge 11   &  0.08  &  -1.76   		 & 18.53 $\pm$ 0.03  & 50.83 $\pm$ 0.81	& 18.54 $\pm$ 0.02	& 51.02 $\pm$ 0.43 	  \\
Reticulum & 0.03  &  -1.57   		 & 18.35 $\pm$ 0.03 & 46.84 $\pm$ 0.68	& 18.42 $\pm$ 0.04 	& 48.35 $\pm$ 0.97   \\
        \hline
    \label{tab:subdwarfs_res}
    \end{tabular}
\begin{tablenotes}[b]
	\item $^a$ E(B-V) adopted from \protect\citet{Walker:1992, Walker:1993}
	\item $^b$ From Table \ref{tab:fehs}.
\end{tablenotes}
\end{threeparttable}
\end{table*}

Radial velocity measurements suggest that even the oldest clusters in the LMC lie in the disk (\citealt{Schommer:1992, Grocholski:2006, Grocholski:2007}). Due to the inclination of the LMC on the sky, a simple average of cluster distances assumes a random distribution and could result in a biased distance estimate. However, if we assume the LMC clusters lie in a disk, geometric variations may be accounted for. The locations of the clusters on the sky are used to correct the individual distances for the inclination of the disk, thereby obtaining a more accurate absolute distance for the LMC. By doing so, we obtain an estimate to the distance to the center of the LMC that is more accurate by accounting for the geometry of the galaxy.

This process is outlined in \cite{van-der-Marel:2001} and \cite{Grocholski:2007}, and also presented in the equations \ref{eq2} through \ref{eq5} below. The distance to the LMC center is denoted D$_0$. D is the distance to any location on the disk plane; for our case, this is the distance to each cluster. We assume the LMC disk to be inclined to the plane of the sky by the angle \emph{i} around an axis at position angle $\theta$. $\theta$ is measured counterclockwise from the west; to align with usual astronomical convention, $\Theta$ is defined the position angle from the north ($\Theta$ = $\theta$ - 90). These values are adopted from \cite{van-der-Marel:2001}; we note that \cite{Grocholski:2007} explored a variety of geometries from other past work, finding differences in the resulting distance of $\lesssim$ 0.2 kpc for varying $\emph{i}$ and $\Theta$. The right ascension and declination are referred to respectively by $\alpha$ and $\delta$ in the equations below.

\begin{equation}\label{eq2}
D/D_0 = \cos{i}/[\cos{i}\cos{\rho} - \sin{i}\sin{\rho}\sin{(\phi - \theta)}]
\end{equation}

To calculate the distance to the LMC center (D$_{0}$ above), it is necessary to calculate $\rho$ and $\sin{(\phi - \theta)}$. The former is calculated as in equation \ref{eq3}, below:

\begin{equation}\label{eq3}
\cos{\rho} = \cos{\delta} \cos{\delta_0} \cos{(\alpha - \alpha_0)} + \sin{\delta} \sin{\delta_0}
\end{equation}

\noindent Via a typical trigonometric expansion, $\sin{(\phi - \theta)}$ may be determined as $\sin{(\phi - \theta)}$ = $\sin{(\phi )}$ $\cos{(\theta )}$ - $\cos{(\phi )}$ $\sin{(\theta )}$. These components are detailed in equations \ref{eq4} and \ref{eq5}:

\begin{equation}\label{eq4}
\sin{\rho} \cos{\phi} = - \cos{\delta} \sin{(\alpha - \alpha_0)}
\end{equation}

\begin{equation}\label{eq5}
\sin{\rho} \sin{\phi} = \sin{\delta} \cos{\delta_0} - \cos{\delta} \sin{\delta_0} \cos{\alpha - \alpha_0}
\end{equation}

Through these equations, we determine an average, error-weighted distance modulus to the LMC center of D$_{0}$ $=$ 18.20 $\pm$ 0.06, equivalent to 46.42 $\pm$ 0.45 kpc. However, NGC 1841 is a clear outlier in these data, lying more than 3-$\sigma$ from the mean D$_{0}$ distance to the LMC center, leading to several possibilities. One interpretation of these results suggests that NGC 1841 may not be a member of the LMC disk or may have been pulled out of the LMC disk (\citealt{Grocholski:2007}). Another view is that an uncertain estimate of the reddening could be affecting the distance measurement; if the reddening of NGC 1841 is doubled from E(B--V) = 0.18, it is brought into agreement with the LMC mean distance. Regardless of the reason(s) for the discrepancy, we remove NGC 1841 as an anomaly and continue determining the distance to the LMC center.

\begin{figure}
  \centering
    \includegraphics[width=0.5\textwidth]{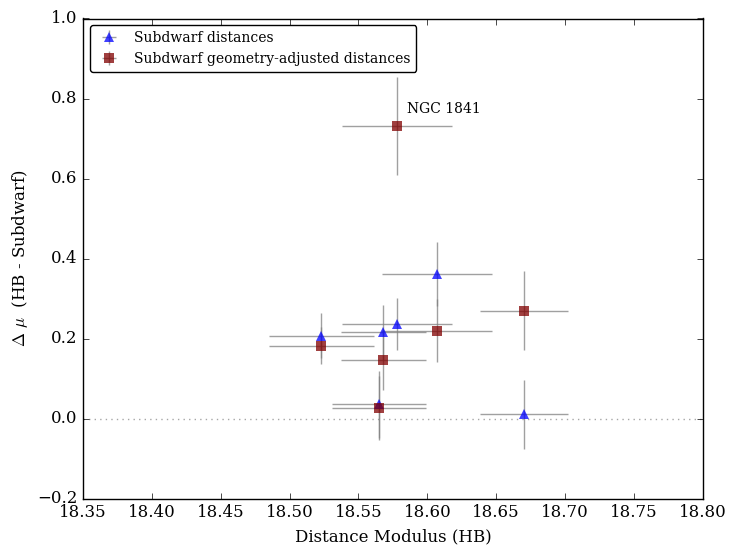}
  \caption{Distances derived from the horizontal branch analysis are compared to the subdwarf distances (blue triangles) and subdwarf distances adjusted for the LMC geometry (red squares). The y-axis is calculated by subtracting the subdwarf derived distances from the horizontal branch distance estimates. The outlying point is NGC 1841. While there is an offset between the two approaches, it appears to be zero-point related and does not vary with distance.}
  \label{fig:distcomp}
\end{figure}

Leaving out the extreme outlier of NGC 1841, the error-weighted mean is \LMCdistDepr (\LMCdistDeprKpc). This is within the range of the result from Section \ref{sec:zahb} from the HB analysis, as seen in Figure \ref{fig:distcomp}, where the subdwarf distances and the subdwarf distances adjusted for the LMC geometry are compared to those derived from the HB analysis. The subdwarf distances tend to be shorter by 0.14 mag on average than those of the HB analysis; the adjusted subdwarf distances are also shorter than the HB distances by 0.12 (leaving NGC 1841 out). The offset appears to be consistent regardless of the individual cluster. The systematic bias between the derived distances could be due to the combination of necessary assumptions in the two approaches (e.g.: metallicity, reddening, GGC distances) or the zeropoints. The HB distance calibration relies on the assumed metallicity-magnitude scale from \cite{Harris:2010}, which is more likely to result in a systematic offset relative to the more fundamentally derived subdwarf distances. Regardless, the overall distance estimates of the LMC cluster system from the two methods are in reasonable (though not excellent) agreement.

Our estimated distance D$_{0}$ to the center of the LMC is also comparable to, though marginally shorter than, recent determinations of the LMC distance of 18.49 $\pm$ 0.09 (\citealt{de-Grijs:2014} and references therein) and  18.493 mag ± 0.008(statistical) ± 0.047 (systematic) (\citealt{Pietrzynski:2013}). Our result is comparable to the values of 18.40 $\pm$ 0.04 and 47.9 $\pm$ 0.9 kpc, derived from \cite{Grocholski:2007} via red clump distances from K-band observations for 25 clusters in the LMC, also taking into account the geometry of the LMC disk. The two studies agree despite differences from our sample being smaller and our focus on the old globular clusters. The true distance modulus derived from \cite{Johnson:1999} of 18.46 $\pm$ 0.09 from three LMC clusters is also comparable to our derived \LMCdistDepr (\LMCdistDeprKpc). In all cases, the results are consistent within the uncertainties.


\section{Conclusions}\label{conclusions}

In this paper, we have used new deep homogeneous photometry for six ancient globular clusters in the Large Magellanic Cloud to measure the epoch of metal-poor globular cluster formation relative to that in the Milky Way, and calculate a new distance estimate to the LMC. Our main results are:

(i) By utilising the same methodology as employed by MF09, we have calculated reddening and distance independent relative ages for our six targets with a mean precision of 8.4\%, and down to 6\% for individual clusters. The LMC sample is coeval with the metal-poor inner halo Galactic GCs measured by MF09 to within 0.2 $\pm$ 0.4 (1.0 Gyr stddev). This is despite the fact that the LMC was most likely widely separated from the Milky Way at this time, and despite the significantly different formation environments implied by the different halo masses of the LMC and Milky Way.

(ii) Through HB magnitude and sub-dwarf fitting methods, we have determined two distance estimates to the LMC. These are \LMCdistHB and \LMCdistSD, respectively. There is a systematic mean difference between the individual cluster distance determinations of $\sim$ 0.14 mag; this could be due to differences in the underlying assumptions for the two different methodologies (for example, the assumed distances of the reference GGCs in the HB analysis vs the assumed metallicities of the LMC clusters during the subdwarf analysis). The average distance from the two methods is \LMCavgdist  (\LMCavgdistKpc), which is comparable to the distance modulus derived recently by de Grijs et al. 2014. 

(iii) To counter the possibility that our distance estimate could be biased if the target clusters are members of the LMC disk, we calculate a geometrically corrected sub-dwarf distance to the LMC of \LMCdistDepr (\LMCdistDeprKpc), where the measurement for NGC 1841 has been excluded as an outlier.


\section*{Acknowledgments}

We thank an anonymous referee whose comments and suggestions were very helpful. D.M. is grateful for support from an Australian Research Council Future Fellowship (FT160100206). D.G. gratefully acknowledges support from the Chilean BASAL Centro de Excelencia en Astrof\'isica y Tecnolog\'ias Afines (CATA) grant PFB-06/2007.

\bibliographystyle{mn2e}
\bibliography{LMC}
\clearpage

\label{lastpage}

\end{document}